\documentstyle[12pt]{article}
\input epsf
\newlength{\extraspace}
\newlength{\extraspaces}
\newcommand{\be}{\begin{equation}}
\newcommand{\ee}{\end{equation}}
\newcommand{\bea}{\begin{eqnarray}}
\newcommand{\nn}{\nonumber}
\newcommand{\eea}{\end{eqnarray}}
\newcommand{\nd}[1]{/\hspace{-0.6em} #1}
\newcommand{\nk}{\noindent}

\def\lsim{\mathrel{\rlap {\raise.5ex\hbox{$ < $}}
{\lower.5ex\hbox{$\sim$}}}}

\newcommand{\pr}{\paragraph{}}
\def\gappeq{\mathrel{\rlap {\raise.5ex\hbox{$>$}}
{\lower.5ex\hbox{$\sim$}}}}
\def\lappeq{\mathrel{\rlap{\raise.5ex\hbox{$<$}}
{\lower.5ex\hbox{$\sim$}}}}

\begin{document}

\begin{titlepage}

\begin{flushright}
OUTP-96-76P \\

hep-th/9701087 \\
\end{flushright}

\begin{centering}
\vspace{.05in}
{\Large {\bf Inverse Landau-Khalatnikov
Transformation and Infrared Critical Exponents
of (2+1)-dimensional Quantum Electrodynamics \\}}
 
\vspace{.2in}
I.J.R. Aitchison, 
N.E. Mavromatos$^{*}$ 
and D. Mc Neill\\
\vspace{.05in}
University of Oxford, Theoretical Physics, 
1 Keble Road OX1 3NP, U.K. \\

\vspace{.2in}
{\bf Abstract} \\
\vspace{.05in}
\end{centering}
{\small By applying an inverse Landau-Khalatnikov 
transformation, connecting (resummed) Schwinger-Dyson 
treatments in non-local and Landau gauges of $QED_3$,
we derive the infrared behaviour 
of the wave-function renormalization in the Landau gauge,
and the associated critical exponents in the normal 
phase of the theory (no mass generation). The result agrees
with the one conjectured in earlier treatments. The
analysis involves an approximation, namely an expansion
of the non-local gauge in powers of momenta in the infrared.
This approximation is tested
by reproducing the critical number of flavours necessary
for dynamical mass generation 
in the chiral-symmetry-broken phase of $QED_3$.}

\vspace{2.5in}

\begin{flushleft} 
January 1997 \\
$^{(*)}$~P.P.A.R.C. Advanced Fellow. \\
\end{flushleft} 

\end{titlepage}

\pr
In recent years interest in three- (space-time) dimensional 
gauge theories has been revived
due to their possible connection with novel 
mechanisms of superconductivity, associated with the recently
discovered high-temperature oxides~\cite{bednortz}.
In the beginning, physicists concentrated their efforts
on understanding the superconducting pairing mechanisms;
later on it was realized that these materials
exhibited unconventional properties even in their normal phase~\cite{normal}.
Such properties suggested deviations from the fermi-liquid behaviour
of the cuprates in their normal phase, and much effort has been 
devoted in the past few years to understanding the physics of the normal
phase. 
\pr
Recently, Landau's fermi liquid theory was reformulated in terms 
of the renormalization group, according to which 
Landau's fermi liquid corresponds to a theory
with a trivial infrared fixed point, i.e. there are no relevant or 
marginal interactions that can drive the theory to a non-trivial
fixed point~\cite{shankar}. 
In this spirit, the observed deviations from fermi liquid behaviour 
in the normal phase of the high-$T_C$ cuprates
might be explained by the presence of a non-trivial 
infrared fixed point in the effective low-energy 
theory describing the physics of the normal state of these materials.
There have been works~\cite{polchinski}
which aimed at studying the interaction of the statistical gauge
field ( which simulates the spin interactions in the magnetic scenaria
for superconductivity~\cite{fetter}) 
with the spinon degrees of freedom, in the context of the
spin-charge
separated phase of the model~\cite{anderson}. 
It has then been argued that this interaction  
is marginally relevant, leading to
a non-trivial
fixed point at low energies.
In ref. \cite{aitmav} it has been argued that the holon-gauge 
interaction is also marginally relevant, leading to a non-trivial 
infrared fixed point. This interaction was also responsible for the phenomenon
of dynamical opening of an energy gap in the holon spectrum, and
thus for superconductivity~\cite{dorey}. 
\pr
The model we studied in ref. \cite{aitmav}
was a standard model of three-dimensional
quantum electrodynamics, with $N$ flavours of fermion fields,
representing the holons. The number $N$, though, was associated
with the angular size of the cells in which we divided 
the fermi surface of the statistical model, 
taken to be a circle of radius $p_F$ 
for simplicity. 
The continuum theory was obtained by linearizing about points
on the surface, the linearization being done by the introduction
of quasiparticles as defined in ref. \cite{benfatto}.
The concept of the quasiparticles was essential in yielding the correct
scaling properties to be used in the renormalization
group approach~\cite{benfatto,aitmav}. 
If $\Lambda$ is an ultraviolet cut-off in
momenta measured above the fermi surface, then the size of the 
angular cell may be taken to be~\cite{aitmav}
\be
      {\rm size~of~angular~cell} \sim \left(\frac{\Lambda}{p_F}\right)^\gamma
=\left(\frac{\Lambda}{p}\right)^\gamma
\left(\frac{p}{p_F}\right)^\gamma 
\label{size}
\ee
where $p \equiv |{\underline p}|$ is the magnitude of the three-vector $p_\mu$,
and $\gamma$ is some (infrared) critical exponent 
to be determined by the renormalization group. 
The momentum scale $p$ is taken to be $p << p_F$.
Infrared physics is studied by taking the 
ultraviolet cut-off $\Lambda \sim p << p_F$. 
The identification of (\ref{size}) 
with the inverse of flavour number $1/N$ implies the 
existence of a running flavour number, i.e. a `running' in theory
space
of three dimensional gauge theories. 
\pr
The cell argument given above seems to be too closely tied to the 
existence of a fermi surface. In ref. \cite{aitmav} we have 
tried to ask a similar question in a relativistic theory,
namely standard quantum electrodynamics 
in three space-time 
dimensions ($QED_3$), where there is no fermi radius. 
\pr
A pleasant 
outcome of the analysis of ref. \cite{aitmav} was 
the confirmation of the running of the flavour number
arising as a consistent solution of the system of 
 Schwinger-Dyson (SD) equations.  
This allows a universal treatment of the running
of the flavour number. The basic reason why the flavour number
in $QED_3$ `runs' (or actually `walks' due to slow running~\cite{aitmav})
is the fact that the standard coupling constant 
of $QED_3$ has dimensions of $\sqrt{mass}$. The ultraviolet (UV)
behaviour of the theory is trivial, $QED_3$ 
is a {\it superrenormalizable} theory, and no UV infinities 
are encountered. However, the infrared (IR) behaviour 
is non-trivial, since dynamical mass generation
occurs at very low energies~\cite{app}. Moreover it is
known that the latter phenomenon occurs only 
below a certain `critical' number of flavours~\cite{app,maris}. 
This already suggests a r\^ole for the flavour number 
as a dimensionless coupling constant of $QED_3$, in which case
the existence of a critical number of flavours would appear as a 
critical value of the coupling, above which dynamical 
mass generation (and hence chiral symmetry breaking) occurs
as a strong coupling phenomenon. 
\pr
Indeed, this interpretation is supported by
large-N treatments of SD equations, where the 
flavour number appears as a dimensionless
ratio of a bare coupling constant $e^2$ and a 
dynamically appearing mass scale $\alpha$.
The latter scale was originally defined by demanding 
$\alpha$ finite as $N \rightarrow \infty$, as in 
standard large-N treatments. 
However, it was immediately realized~\cite{app} that 
all the momentum loop integrals of $QED_3$ in a SD analysis
were damped quickly for momentum scales 
higher than $\alpha$, and hence the latter was considered
as 
a spontaneously appearing 
mass scale in the problem playing the r\^ole of 
an effective UV cut-off. 
The main point of ref. \cite{aitmav} was then to show
that there exists a slow running of the `renormalized' 
$e^2$, at intermediate scales of momenta. In terms of the 
statistical model, the scale $\alpha$ may be identified
with $p_F$. In such a case, 
the infrared physics is equivalent to a large-N
approximation, naturally induced in the problem~\cite{shankar,aitmav}.
If there is a non-trivial infrared fixed point, then, 
the exponent $\gamma$ behaves as a critical exponent
characteristic of the model. Indeed, the analysis 
of ref. \cite{aitmav} has shown that the scaling (\ref{size})
also characterizes the wave-function renormalization 
of the SD equations for dynamical mass generation 
in $QED_3$, while 
the effective 
static potential behaves as 
\be 
    V(p) \sim p^{-1 + 2\gamma} 
\label{effpot}
\ee
and, thus, the critical exponent $\gamma$ 
describes the deviation from the Coulombic
potential,
and therefore from 
the Landau fermi-liquid (trivial) fixed point~\cite{polchinski,aitmav}.
\pr
A different set of exponents also characterizes 
the finite-temperature scaling of the resistivity 
of the theory, the latter being defined as the response 
of the system to an externally applied (electric)
field. In this article we shall not discuss
the finite-temperature analysis. Some preliminary
discussion of this important topic may be found in 
ref. \cite{aitmav}, where the temperature is viewed as 
a sort of finite-size (infrared cut-off) scaling in the problem. 
\pr
The renormalization of the dimensionful coupling, $e^2$,  
and the interpretation of $\alpha$ as the effective cut-off scale, 
are supported by a Wilsonian approach to the Renormalization Group
(RG),
where the running of a coupling is a consequence of 
integrating out degrees of freedom, not necessarily 
associated with UV divergences. 
However, in our problem, as we 
discussed in ref. \cite{aitmav}, there is an infrared singular 
behaviour, and in
some 
sense the above-mentioned
slow running is associated with it.
It is the purpose of this article to elaborate further
on this latter point. 
\pr 
The running of $e^2$ and the interpetation of $\alpha$ as a 
cut-off scale in the Wilsonian RG approach, imply an 
effective running of the dimensionless coupling constant
$g$ of the problem, i.e. the effective flavour number~\cite{aitmav}:
\be
     g_0 \equiv \frac{1}{8\pi^2N}  
\rightarrow g_R = \frac{1}{8\pi^2N(p/\alpha)}  
\label{flavour}
\ee
with $p$ a momentum scale. 
\pr
The analysis of ref. \cite{aitmav} was carried out 
in the Landau gauge and to leading order in $1/N$ expansion. The 
system of SD equations in this case reads
\bea 
&~&A(p) = 1 -\frac{\alpha}{\pi^2N}\frac{1}{p^3}
\int _0^\infty  dk \frac{k A(k)G(p^2, k^2)}{k^2A^2(k)+B^2(k)}
I(p,k) \nn \\
&~&B(p) = \frac{\alpha}{\pi^2 N p} \int _0^\infty
\frac{kB(k)G(p^2, k^2) }{k^2A^2(k) + B^2(k) } 
\{ 4{\rm ln}\frac{ p+k + \alpha}{|p-k| + \alpha} \} \nn \\
&~&I(p,k) \equiv \alpha^2 {\rm ln}\frac{p+k + \alpha}{|p-k|
+ \alpha} - \alpha (p + k - |p-k| + 2pk - 
\frac{1}{\alpha}|p^2 - k^2| (p+k - |p-k|) - \nn \\
&~&\frac{1}{\alpha ^2}
(p^2 - k^2)^2 \{ {\rm ln}\frac{p+ k + \alpha}{|p-k| + \alpha} 
-{\rm ln} \frac{p+k}{|p-k|} \} 
\label{sdeq}
\eea
where $A(p)$ is the wave-function renormalization, $B(p)$ 
is the gap function, and  
the vertex function is $\Gamma _\mu = \gamma _\mu G(p^2, k^2) $. 
In ref. \cite{aitmav},
we  made
a simple vertex choice $\Gamma _\mu = A(p)\gamma _\mu$, consistent 
with the Ward-Takahashi identities~\cite{kn,penn} stemming from 
gauge invariance.
\pr
The point of the analysis of ref. \cite{aitmav} was to look for 
consistent solutions of (\ref{sdeq}) in the normal phase of 
the model, where there is no dynamical mass generation. 
This allows a solution of the wave-function renormalization 
equation in the SD system, by the bifurcation method~\cite{kn},
i.e. setting $B(p)=0$ in the denominators. Substituting, then, the 
solution for $A(p)$ into the equation for the mass gap $B(p)$ one may
define an effective `running coupling'~\cite{higashijima,kn,aitmav}
\be
   g_R \equiv \frac{g_0}{A(p)} 
\label{running}
\ee
The main result of \cite{aitmav}, which improved and 
extended considerably an 
earlier treatment in ref. \cite{kn} in the dynamical mass
generation
regime, was to show that (\ref{running}) 
runs with the momentum scale $p/\alpha$ (which  plays the r\^ole 
of an effective RG scale) as in fig. 1.
The introduction of an infrared (IR) cut-off scale~\cite{kn,aitmav}
leads to the appearance of a `fixed-point' structure in the infrared.
Removal of the cutoff is a subtle issue and depends
on the type of IR cut-off used~\cite{aitmav}. 
\pr
Subsequent to the work of \cite{aitmav}, an analysis 
has been performed~\cite{kondo} in the so-called non-local
gauge~\cite{nlg},
confirming the results of \cite{aitmav} on the slow running 
of the coupling at intermediate momentum scales $k \lsim \alpha$, 
as compared to earlier approximations
in a SD analysis of $QED_3$~\cite{kn}, as well as on the 
existence of an infrared fixed point. The running coupling in the non-
local gauge is again defined through the mass gap equation of the SD
system, and it turns out to be related to the Fourier transform of the 
non-local gauge itself~\cite{kondo}:
\be
   g_R^{nlg} (t)/g_0^{nlg} = 1 + \frac{1}{2}{\tilde \xi}(k^2)
\qquad ; \qquad t = {\rm ln}(k/\alpha)
\label{nlgft}
\ee
An important feature of the non-local gauge is that 
in this gauge the wave-function 
renormalization is equal to one identically:
\be
    A(p) \equiv 1 
\label{wfnlg}
\ee
Therefore, one would  
like to study the formal connection of this result 
to the one derived in the Landau gauge, where the 
deviation of the wave function renormalization from 1
lead to the running coupling (\ref{running}) (c.f. fig. 1). 
In ref. \cite{kondo} it has been remarked that such a connection
is provided by the so-called Landau-Khalatnikov (LK) 
transformation~\cite{nlg,konebi}, which by construction 
leaves the system of SD equations {\it form invariant}. 
\pr
It is the purpose of this short note to apply the (inverse) 
LK transformation and determine the behaviour of the 
wave-function renormalization in the infrared regime of the theory
in the Landau gauge, starting from quantities
evaluated in the non-local gauge. As we shall show, 
in the absence of an infrared cutoff, where we shall
restrict our attention for simplicity, 
the low-momentum behaviour of the 
Landau-gauge wave function renormalization is
\be
   A(p) \sim \left( \frac{p}{\alpha} \right)^{\gamma} 
\qquad \gamma =\frac{8}{3N\pi^2} 
\label{irwf}
\ee
with $\alpha$ playing the r\^ole of an effective UV cut-off.
The exponent $\gamma$ appears as a sort of critical exponent
characterizing the behaviour of the theory in the infrared. 
The result (\ref{irwf}) has been conjectured in the past,
based on either theoretical speculations~\cite{app2,penn,atk},
or in preliminary numerical treatments~\cite{maris}.
The exponent can also be derived from
a qualitative study of the  
renormalization-group running of the effective coupling
in the infrared within a SD treatment~\cite{aitmav}. 
In this article we derive 
this exponent {\it analytically} by
making a leading-order (infrared) approximation in the 
LK transformation 
connecting results in the Landau gauge with 
those in the non-local gauge.
To our knowledge this 
is the first analytic derivation of a critical exponent
in $QED_3$. The presence of this exponent complements 
the results of refs. \cite{aitmav,kondo} about the existence of a  
non-trivial IR fixed point, 
and, in view of our statistical interpretation in (\ref{size}),
(\ref{effpot}),  
determines the universality class of the  
deviations from the fermi-liquid (trivial) fixed point
characterizing 
superconductor models based on $QED_3$~\cite{dorey,aitmav}. 
\pr
We now proceed to the derivation of $\gamma$. 
Our starting point will be the LK transformation
between the non-local and Landau gauges in the 
normal phase (no dynamical-mass generation). 
The fermion propagator in the non local gauge
in momentum space reads~\footnote{We follow 
the convention: $\{ \gamma _\mu, \gamma _\nu \} =
-\delta _{\mu\nu} $,
$\nd{p}\nd{p}=-p^2$.}:
\be 
  S'_F(p)=S_{F,(0)}'(p)=1/\nd{p}
\label{fpnlg}
\ee
which implies that  in configuration space 
\be
  {\tilde S}_{F,(0)}'(x) =-\int \frac{d^3p}{(2\pi)^3}
e^{-ip_\mu x^\mu}\frac{\nd{p}}{p^2}=i\nd{x}{\cal P}_0(x)
\qquad ; \qquad {\cal P}_0(x) \equiv \frac{\Gamma (3/2)}{2\pi^{3/2}
|x|^3}
\label{conffp}
\ee
In Landau gauge, on the other hand, 
the fermion propagator reads:
\be
S_F(p)^{-1}=A(p) \nd{p}
\label{lgfp}
\ee
{}From the form invariance of the SD equation in the two gauges
one may obtain a formal expression of the wave function 
renormalization in terms of the non-local gauge~\cite{nlg,kondo}:
\bea
&~&   A^{-1}(p) = -i \int d^3xe^{ip_\mu x^\mu}
e^{-\Delta (x)} p_\mu x^\mu {\cal P}_0 (x) \nn \\
&~& \Delta (x) \equiv e^2 \int \frac{d^3k}{(2\pi)^3} 
(e^{-ik_\mu x^\mu} - 1) F(k) ~;\qquad F(k) = \frac{{\tilde \xi}(k^2)}{k^4
(1 - \Pi (k^2) /k^2)}
\label{nlgwfr}
\eea
where ${\tilde \xi}(k^2)$ is related to the Fourier transform 
$\xi (k^2)$ of the non-local gauge by:
\be
   {\tilde \xi}(k^2) \equiv \xi (k^2) (1 - \Pi (k^2)/k^2)
\label{defxtilde}
\ee
and $\Pi (k^2)$ is the photon vacuum polarization. In the non-local gauge,
in the resummed $1/N$ SD treatment, there is an {\it exact}
expression for $\Pi (k^2)$ which is given by~\footnote{The exactness
of the photon polarization is a consequence of the fact that
in the non-local gauge 
$A(p) \equiv 1$, and the 
vertex function, consistent with Ward-Takahashi identities
stemming from gauge invariance, may be taken 
to be the trivial one, $\Gamma _\mu = \gamma _\mu$~\cite{kondo}.}:
\be 
  \Pi (k^2) = -\alpha k \qquad ; \qquad k \equiv |{\underline k}|
\label{photonpol}
\ee
Thus, the function $F(k)$ is known, once we know the non-local gauge 
${\tilde \xi}(k^2)$. From the analysis of 
ref. \cite{kondo} we have 
\be
{\tilde \xi}(k^2) =1 + \frac{1}{(k^2)^2 D_T (k^2)} \int _0^{k^2}
dz \frac{d}{d z } D_T (z) z^2 
\label{expre}
\ee
with the transverse photon propagator $D_T$ given by
\be
   D_T (k^2) = \frac{1}{k^2 + \alpha k} \qquad, \qquad D_T (z)
=\frac{1}{z + \alpha \sqrt{z}} 
\label{trans}
\ee
as a result of (\ref{photonpol}). 
A straightforward computation yields: 
\be
{\tilde \xi}(k^2) = 2 - 2 \frac{k^2+ \alpha k}{k^2} \left(
1 - \frac{2\alpha}{k} + \frac{2\alpha ^2}{k^2}{\rm ln}(1 +
\frac{k}{\alpha}) \right) 
\label{log}
\ee
and, thus, 
\be
   F(k) = \frac{2}{k^2 (k^2 + \alpha k)}- 
\frac{2}{k^4} \left(1 - \frac{2\alpha}{k} + \frac{2\alpha ^2}{k^2} 
{\rm ln}(1 + \frac{k}{\alpha}) \right) 
\label{fk}
\ee
In the infrared regime, $k << \alpha$, 
where our interest 
lies, one may expand (\ref{fk}) in powers of $k/\alpha$: 
\be
   F(k) \simeq \frac{2}{3\alpha} \frac{1}{k^3} - \frac{1}{\alpha^2 k^2} 
+ \frac{6}{5 \alpha ^3 k} \qquad k \rightarrow 0
\label{fklim}
\ee
We wish to evaluate the Fourier transform of (\ref{fklim}):
\bea
  &~& I \equiv e^2F(x)= 
 e^2\int \frac{d^3k}{(2\pi)^3}e^{-ik_\mu x^\mu}
[ \frac{2}{3\alpha}
\frac{1}{k^3} - \frac{1}{\alpha ^2 k^2} 
+ \frac{6}{5\alpha^3 k} ] = \nn \\
&~&\frac{8}{N}\int
\frac{d^3k}{(2\pi)^3}e^{-ik_\mu x^\mu} [\frac{2}{3k^3} 
- \frac{1}{\alpha k^2} + \frac{6}{5\alpha^2 k} ] 
\qquad ; \qquad  
e^2 \equiv \frac{8\alpha}{N}
\label{ft}
\eea
The integral needs regularization in both the ultraviolet and 
infrared regimes. We choose to regularize 
the UV infinities by dimensional regularization, introducing the 
standard RG scale $\mu$.
Using~\cite{gelfand}:
\be
    \int \frac{d^3k}{(2\pi)^3}e^{-ik_\mu x^\mu} \frac{1}{k^\lambda}
= \frac{1}{2^{\lambda}\pi^{3/2}}\frac{\Gamma (\frac{3-\lambda}{2})
|x|^{\lambda-3}}{\Gamma(\frac{\lambda}{2})}
\label{gelf}
\ee
one arrives at:
\be
 \frac{16}{3N \mu^{d-3}}\int
\frac{d^dk}{(2\pi)^d k^3}e^{-ik_\mu x^\mu}  
= \frac{16}{3N \mu^{d-3}}\left(\frac{2^{d-3}\pi^{d/2}}{(2\pi)^d}
\frac{\Gamma (\frac{d-3}{2})}{\Gamma (3/2)}|x|^{3-d}\right)
\label{regul}
\ee
which in the limit $d-3 =\epsilon \rightarrow 0^+$ yields:
\be
  I = \frac{24}{5N\pi^2}\frac{1}{|\alpha x|^2}
- \frac{2}{N\pi}\frac{1}{|\alpha x|} + \frac{4}{3N\pi^2}
\left(\frac{2}{\epsilon} + 2 {\rm ln}\frac{1}{|\mu x|} - \gamma_0 \right)
\label{result}
\ee
where $\gamma _0$ is the Euler-Macheroni constant. 
\pr
Next we have to evaluate $e^2 F(0)$, which enters
the definition of $\Delta (x)$ appearing in (\ref{nlgwfr}),
as a result of the requirement that the coincidence
limit of the free fermion propagator is
the same in all gauges~\cite{nlg,konebi}.
To this end, one needs to regularize the coincidence limit $e^2F(0)$. 
We do so by replacing the argument by $\hbar c/\alpha$, with $\alpha$
viewed as our effective UV cut-off scale 
(in the system of units we are working, we set 
$\hbar c =1$). 
Then we have:
\be
\Delta (x) = e^2 (F(x) - F(1/\alpha)) \sim 
\frac{24}{5N\pi^2\alpha^2}(\frac{1}{x^2} - \alpha^2)-
\frac{2}{N\pi \alpha} (\frac{1}{|x|} - \alpha ) + \frac{8}{3N\pi^2} 
{\rm ln}\frac{1}{|x\alpha|} 
\label{dx}
\ee
Notice that the UV regularization mass scale $\mu$ disappeared
from (\ref{dx}). This is a consequence of the fact that
$QED_3$ is super-renormalizable in the UV, so only infrared running 
should emerge, as we shall verify immediately below. 
\pr
Indeed, we can now give an estimate 
of the wave-function renormalization in the infrared regime $p
\rightarrow 0$:
\bea
&~&A^{-1}(p) \sim -i\int d^3x e^{ip_\mu x^\mu} p_\mu x^\mu 
{\cal P}_0(x) e^{-\Delta (x)} \sim \nn \\
&~&-(\alpha)^{\frac{8}{3\pi^2N}} e^{ \frac{24}{5N\pi^2}
- \frac{2}{N\pi}}\frac{p_\mu \nabla _{p_\mu}}{4\pi}
\int \frac{d^3 x}{x^{3-\frac{8}{3N\pi^2}}}e^{ip_\mu x^\mu}  
e^{ - \frac{24}{5N\pi^2\alpha^2 x^2} + \frac{2}{N\pi\alpha |x|} }
\label{ap}
\eea
Since $|x|$, in our regularized coincidence limit, 
is not allowed to go below $1/\alpha$, the factors ${\rm exp}\{-{\cal
O}(\frac{1}{x^2}) + {\cal O}(\frac{1}{|x|}) \}$ are of order one, and
the above analysis yields as $p \rightarrow 0$:
\be
 A^{-1}(p) \sim
e^{\frac{24}{5N\pi^2}-\frac{2}{N\pi}}
\frac{2^\gamma \pi^{\frac{3}{2}}\gamma \Gamma(\frac{\gamma}{2})}{4
\pi \Gamma(\frac{3-\gamma}{2})}\left(\frac{\alpha}{p}\right)^\gamma 
\qquad ; \qquad \gamma = \frac{8}{3N\pi^2} 
\label{critexpo}
\ee
The prefactor in (\ref{critexpo}) depends on the (bare) flavour
number.
To get an order of magnitude estimate, we note that for $N \ge 5$
(where dynamical mass generation does not occur, c.f. below) this
prefactor
is of ${\cal O}(1)$: 
\be
A^{-1}(p) \sim 
0.993\left(\frac{\alpha}{p}\right)^\gamma 
\qquad ; \qquad N=5
\label{ce2}
\ee
This result has been conjectured~\cite{app,penn} but here we proved it
analytically, starting from the non-local gauge. 
\pr
The reader might worry that our result has been obtained on the basis of
the approximation (\ref{fklim}),which arises from a momentum expansion 
in the infrared $k << \alpha$. 
To substantiate its validity we next perform 
an analysis of  dynamical mass generation in the non-local gauge, 
upon using (\ref{fklim}). As we shall show, in this way one 
can reproduce very simply
 the result on the existence of a critical number 
of flavours, which has been obtained earlier based on more exact  
 treatments~\cite{maris,konebi}. 
\pr
We start our analysis from the expression on the gap function 
$B(p)$ in the non-local gauge~\footnote{In this gauge $B(p)$ 
is identical 
to the mass function.}~\cite{kondo}:
\be
B(p) = e^2 \int \frac{d^3q}{(2\pi)^3}\frac{B(q^2)}{q^2 + B(q^2)}
\left( \frac{4}{k^2 + \alpha k } - \frac{2}{k^2}(1 - \frac{2\alpha}{k}
+ \frac{2\alpha ^2}{k^2} {\rm ln}(1 + \frac{k}{\alpha})) \right) 
\label{bp2}
\ee
with $k=\sqrt{q^2 + p^2 - 2pq {\rm cos}\theta}$. 
Expanding the logarithm in the infrared, approximating 
$F(k)$ as 
\be 
k^2F(k) \simeq \frac{2}{k^2 + \alpha k} - \frac{4}{3\alpha k} 
\label{fk2}
\ee
and
using $\alpha$ as an effective UV cut-off of the momentum integrals,
one obtains after performing the angular integrations in (\ref{bp2}):
\be
B(p^2) =\frac{e^2}{4\pi^2} \int_0^\alpha dq B(q) \frac{q^2}{q^2 + B^2 (q)}
\{ \frac{4}{pq} {\rm ln}\frac{p + q + \alpha}{|p-q| + \alpha } 
- \frac{4}{3\alpha pq} (p + q -|p-q|) \} 
\label{bq2}
\ee
Following standard analysis we can approximate~\cite{kn} 
\be
{\rm ln} \frac{p+q+ \alpha}{|p-q|+\alpha} \simeq \frac{2q}{p+\alpha}
\Theta (p-q) + \frac{2p}{q + \alpha} \Theta (q-p)
\label{thetapprox}
\ee
Upon substituting into (\ref{bq2}) we may then convert the integral 
equation into a differential one for low momenta $p$: 
\be
  \frac{d}{dp}B(p^2) = \frac{16}{3N\pi^2}
\frac{p^2 - 4\alpha p - 2\alpha^2}
{p^2(p+\alpha)^2} \int_0^p dq B(q) \frac{q^2}{q^2 + B^2(q)} 
\label{dbdp}
\ee
This differential equation should be solved together with 
the following boundary conditions (consistent with the original 
integral equation)~\cite{app,dorey}
\bea
  &~&  p^2 \frac{d}{dp}B(p)|_{p=0} = 0 \nn \\
&~& \left(B(p) + \frac{4\alpha}{10}\frac{d}{dp}B(p)\right) |_{p=\alpha} =0 
\label{bc}
\eea
The region relevant for dynamical 
mass generation is $p << \alpha$; in this region the differential equation
(\ref{dbdp}) becomes:
\be 
   \frac{d}{dp} \left( p^2 \frac{d}{dp} B(p) \right) =
-\frac{32}{3N\pi^2}
\frac{p^2 B(p) }{p^2 + B(p)^2} 
\label{approxde}
\ee
Following standard treatments~\cite{app,dorey}, then, 
it is immediate to see that
dynamical mass generation occurs only for 
flavour numbers smaller than the following critical value:
\be
  N_c \simeq \frac{128}{3\pi^2} \simeq 4.32
\label{crit}
\ee
The number (\ref{crit}), obtained here from an infrared approximation
in the expression for the non-local gauge, (\ref{fk2}),
agrees with more exact treatments ~\cite{maris,konebi}. 
This agreement offers support for the 
method used here to derive the 
critical exponents (\ref{critexpo}) in the normal phase,
where a similar  infrared approximation, (\ref{fklim}),
 used. Notice also that the 
result (\ref{crit}) 
agrees remarkably with that of an earlier treatment~\cite{nash}, 
incorporating $1/N^2$ corrections to 
the SD equations. 
\pr
The above analysis, then, shows that although 
the arguments of ref. \cite{penn} on the infrared behaviour
(\ref{critexpo}) of the wave-function renormalization
were correct,  they were not sufficient 
to destroy the existence of a critical number of flavours
for dynamical mass generation, within a resummed-$1/N$ 
treatment of the SD equations. The existence of a critical 
number of flavours for  dynamical mass generation implies 
that the renormalization-group diagram of fig. 1 
will describe dynamical mass generation in the infrared 
if and only if the height of the abscissa at the orgin $p=0$
is large enough. 
\pr
It should be noted at this stage that the 
presence of an infrared cut-off, which was ignored in this article, 
will alter the infrared behaviour of the running coupling by 
cutting off its growth as in fig. 1~\cite{aitmav}, and -in the 
case of dynamical mass generation - will induce an 
IR-cut-off-dependent $N_c$~\cite{kn}. 
In such a case, 
one could imagine starting from a 
non-local gauge with a covariant IR cut-off~\cite{kondo}, 
and applying the inverse LK transformation to reproduce the 
results of ref. \cite{aitmav} for the infrared behaviour 
of the wave-function renormalization. 
It should be stressed, 
however, that the critical exponent $\gamma$ acquires physical 
meaning, as a quantity 
characterizing the universality class of the theory
(in standard RG language),
only after removal of the IR cut-off. 
We now note that the RG running of the effective coupling  
in the presence of a covariant IR cut off appears to 
have discontinuities, which make the removal of the cut off 
problematic.
Such problems have not been resolved yet~\cite{aitmav,kondo}, 
but we note that such discontinuities might not be so unphysical,
given their resemblance to  Landau-damping discontinuities appearing in
finite-temperature field theories~\cite{zuk},with which a theory with 
a covariant IR cut off has many things in common~\cite{aitmav}.
We hope to return to a careful study of 
such issues in the near future. 
\pr
The existence of critical exponents, characterizing the infrared
behaviour
of the model, acquires physical significance
when the theory is viewed as a model describing 
the physics of the planar high-$T_c$ materials,
since in that case it characterizes in a universal way the 
deviations from the fermi-liquid fixed point, and thus 
is subject to experimental tests~\cite{normal,shankar}. 
In the present article, 
the critical exponents have been derived in a rather cavalier 
way, i.e. by performing a renormalization procedure based on a 
large-N (resummed) SD analysis. A more exact treatment
can be provided by the application of the exact renormalization-group
approach due to Wilson~\cite{wilson}, 
appropriately adapted  to incorporate
large-N treatments. Such a programme, when performed, will yield
useful information/verification on the infrared behaviour 
of the model, and will determine in a more accurate way
the associated critical exponents. It may also shed some light 
on resolving 
the issue of the discontinuities of the 
theory with a covariant IR cut-off,
found in refs. \cite{aitmav,kondo}. 
Such an analysis 
is left for future work.   
\pr
\nk {\Large {\bf Acknowledgements} }
\pr
We wish to acknowledge useful discussions with M. D'Attanasio, 
K.-I. Kondo and T. Morris, and to thank K.-I.Kondo for a careful 
reading of the manuscript.
D.McN. wishes to thank P.P.A.R.C. (UK) 
for a research studentship. 

\newpage

\newpage

\begin{figure}[htb]
\epsfxsize=3.2in
\centerline{\epsffile{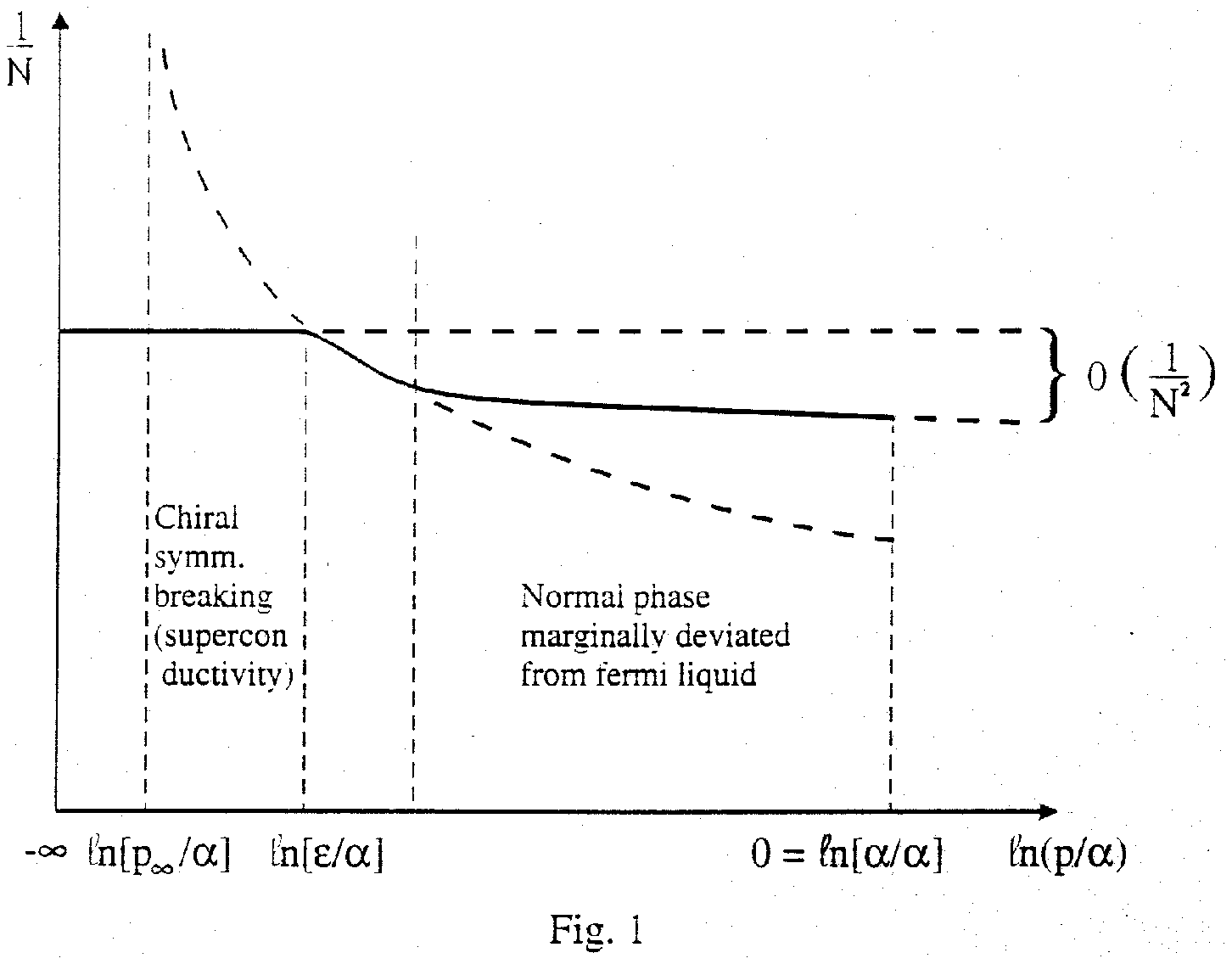}}
\caption{Running flavour number in $QED_3$ versus 
a momentum scale $p$, within a 
resummed-$1/N$ Schwinger-Dyson treatment, in the presence of  
an infrared cut-off $\epsilon$. The running is slow, and there exists 
a non-trivial infrared (IR) fixed point. This kind of behaviour 
is argued to be responsible for (marginal) deviations 
from the fermi-liquid (trivial IR fixed point) theory.
}
\label{fig1}
\end{figure}

\end{document}